**Plasmon-enhanced Hyperspectral Imaging**

*Mr. Kristian Caracciolo, Dr. Eugeniu Balaur\*, Assoc. Prof. Walter D. Fairlie, Assoc. Prof. Erinna F. Lee, Dr. Jacqueline M. Orian, Prof. Eric Hanssen and Prof. Brian Abbey\**

\*E. Balaur, B. Abbey
Plenty Rd & Kingsbury Drive, 3073, Australia
E-mail: E.Balaur@latrobe.edu.au , B.Abbey@latrobe.edu.au

Hyperspectral imaging is gaining attention in the field of disease diagnosis due to its ability to enhance tissue contrast, surpassing the capabilities of conventional brightfield imaging techniques. Typically, histological sections lack sufficient intrinsic contrast in the visible spectrum, necessitating the use of dyes or stains for adequate visualization. However, a recent breakthrough involves the application of plasmonic meta-materials as substrates for histological sections, replacing staining or labelling on traditional microscope glass slides. These nanofabricated microscope slides, shortened to nanoMslides, operate by selectively transmitting colors based on refractive index variations within the sample when illuminated with white light. This study investigates the feasibility of integration of nanoMslides for hyperspectral imaging. By employing a tunable light source, specific plasmon resonances within the slides can be selectively excited. This precise control over plasmonic interactions results in significantly heightened sensitivity and specificity, showcasing the potential for advanced applications in disease diagnosis and biomedical research.

1. Introduction

Hyperspectral imaging (HSI) refers to imaging modalities that capture spatial and spectral information. The spectral information captured is typically either the reflectance or transmission of a sample over the visible to infrared region. HSI systems are broadly separated into spatial and spectral scanning modes. Two common examples of the former are whiskbroom and pushbroom systems that collect spectra pixel by pixel or row by row, respectively, and require spatial scanning of either the source or sample.[1] Spectral scanning modes capture full field images of the sample and either scan the wavelength of the illuminating field or the wavelength detected to resolve spectral information.[1] HSI has seen many applications ranging from forensics to food quality control.[2,3] In these fields the appeal of HSI has been the promise of greater chemical sensitivity without the need to be more invasive than regular optical imaging. Recently, there has been an increasing interest in HSI applications to medical imaging for much the same reason; disease diagnosis that is less invasive or requires less modification of the sample. Medical studies employing HSI have demonstrated applications in the assessment of burn wound severity, evaluation of peritoneal fibrosis, cancer margin detection of pancreatic tissue and classification of ductal carcinoma in situ.[4–7] For many diseases, histology in which a biopsy is taken, sectioned and stained before assessment by a pathologist remains the gold standard for diagnosis.[5,8,9] Thus, studies seeking to apply HSI to disease diagnosis often measure themselves against traditional histology by comparing sensitivity and specificity of the HSI modality to histology often with improvements over traditional histology.[9] Depending on the type of disease, HSI measurements have been demonstrated in vivo and for both stained and unstained biopsy sections.[7,10,11]

Due to their increased dimensionality, data sets obtained from HSI are large and contain some redundant information. To extract their most useful components and reduce their size, hyperspectral data sets, referred to as hypercubes, are often processed. One approach commonly employed is principal component analysis which seeks to eliminate redundancy between different spectral bands and subsequently highlight key features of the sample.[1] Simpler approaches may include projection of the spectral max or mean values to produce a single 2D image. The application of machine learning to the interpretation of hypercubes is becoming increasingly popular. [7,9,10,12]

2. Plasmonic Devices

2.1. Surface plasmon polaritons

Plasmonic devices are a type of metamaterial which exploits the properties of surface plasmon polaritons (SPPs). SPPs are collective oscillations of the free electrons at a metal-dielectric interface excited by electromagnetic radiation. The oscillations are confined to and propagate parallel to the metal/dielectric with an evanescently decaying electric field normal to the surface. SPPs are a long-studied phenomenon first observed in the loss spectra of electron beam diffraction experiments in metals.[13] It was later discovered that SPPs could be excited using light and glass prisms in the Kretschmann and Otto geometries.[13] The dispersion relation for SPPs without damping is shown in **equation 1**, where $k_{sp}$ is the wavevector of the surface plasmon, $k_0$ is the wavector of free light and $\epsilon_d$ and $\epsilon_m$ are the dielectric constants of the dielectric and metal present at the interface, respectively. A consequence of this relation is that for metals in which SPPs are observed, the wavevector of the SPP is greater than that of free light.

$$k_{sp} = k_0 \sqrt{\frac{\epsilon_m \epsilon_d}{\epsilon_m + \epsilon_d}} \qquad (1)$$

SPPs can also be excited by free light if some additional momentum in the plane of the interface is provided to the light by periodic structures.[13] In this case, the wavelength of the SPP mode can be described by **equation 2** where P represents the periodicity of the array and $i$ and $j$ describe the scattering order from the array.[14]

$$\lambda_{sp} = \frac{P}{\sqrt{i^2 + j^2}} \sqrt{\frac{\epsilon_m \epsilon_d}{\epsilon_m + \epsilon_d}} \qquad (2)$$

The periodic structures may be either protrusions from or apertures through the surface as shown in **Figure 1**. If these structures are isolated nanoscale objects, they exhibit a related resonant phenomena known as localised surface plasmons (LSPs). LSPs are analogous to SPPs only that they are confined to the surface of the structure they exist on and do not require the presence of a periodic structure to be excited directly by free light.[13,15] LSPs are responsible for the unique and widely studied light interactions of gold nanoparticles.[16]

**2.2 Extraordinary optical transmission & contrast enhancement**

Extraordinary optical transmission arises from the complex interaction of several resonant phenomena excited in thin metal films bearing apertures with dimensions near or below the diffraction limit. In films with an array of apertures, SPPs are the dominant phenomenon.[13] Traditionally, transmission through an aperture was thought to transition from propagating to evanescent decay as the wavelength of the illuminating light decreases below the cutoff wavelength of that aperture; a combination of the radius of the aperture and the skin depth of the metal.[15] However, when resonant phenomena such as SPPs and LSPs are excited, the electric field above the aperture is enhanced to an extent where the transmission efficiency normalised to the area of the aperture may exceed one, an effect called extraordinary optical transmission (EOT).[15] This work exploits EOT and the extreme sensitivity of SPPs to dielectric constant variations in in the near surface region to greatly enhance sample contrast.

The devices feature 150 nm metal films deposited on glass substrates perforated by nanoscale apertures (see Methods). Two separate device architectures were used in this work; one featuring circular apertures in a symmetric hexagonal array (500 nm periodicity) and one using circular apertures in an asymmetric rectangular array (400 x 500 nm periodicity). The asymmetric aperture spacing allows for two different sets of SPP resonances to be accessed when using linearly polarised light aligned with each spacing as described in equation 2. The terms transverse electric (TE) and transverse magnetic (TM) mode are used to describe images collected using light aligned with the 400 nm and 500 nm direction, respectively.

As the dispersion relation (Equation 1) is directly linked to the dielectric constant in the near surface region of the metal, the wavevector of the SPPs supported at the interface changes when any material is introduced to the surface. Consequently, the intensity of the transmitted light at a particular wavelength in that region of the device changes. When using a brightfield source, this translates to a pronounced colour change. This is in stark contrast to conventional brightfield imaging (Figure 1a) where the interaction of the incident light with transparent objects leads to only a slight decrease in the overall intensity of the transmitted light. When a sample like a histological section is deposited on a plasmonic device, variations in the dielectric constant of the tissue highlight different regions in striking colours without the need for stains. This is demonstrated in Figure. 1b, c for ultrathin sections of optic nerve tissue that are almost completely transparent when collected on a traditional glass slide (Figure 1b right) compared to plasmonic slides (Figure 1c right). The sensitivity of the resonance condition of the device to changes in dielectric constant is great enough that differences between healthy and invasive breast tissue cells has been demonstrated as an alternative to common biomarkers/stains as H&E and KI67.[17]

**2.3 Plasmon-enhanced hyperspectral imaging**
The aim of this work is to demonstrate the feasibility and appeal of combining hyperspectral imaging with plasmonic devices using the setup represented in Figure 1d. Both HSI and the plasmonic slides have demonstrated the ability to enhance chemical sensitivity in isolation.[5,17,18] Although brightfield imaging using nanoMslides shows a dramatic increase in the overall contrast (Figure 1c right) compared to the brightfield imaging using standard microscope slides (Figure 1b right), the color of features in the resulting image is a projection of all wavelengths.  However, by incorporating an HSI system, this limitation can be overcome as spectral plasmonic contrast can be probed for each wavelength (Figure 1d right).

We call this type of imaging plasmon enhanced HSI (PE-HSI). When illuminating the plasmonic device with light of a specific wavelength, specific plasmonic resonances are excited resulting in dramatic contrast changes between wavelengths. Additionally, combining hyperspectral imaging with a polarisation sensitive plasmonic device allows large changes in contrast to be obtained at a single wavelength. The combination of plasmonics and HSI has not been previously explored in the context of histology. Some work has been done applying HSI to gold nanoparticles in solution wherein the spectra of the particles are modified by changes to their local environment.[19] However, this method images only the nanoparticles themselves and would be unsuitable to imaging entire tissue structures. Other works exist imaging the hyperspectral characteristics of plasmonic nanostructures.[20] However, these works characterise the spectral properties of the device itself without investigating any sample and used an electron beam as the means of exciting surface plasmons.

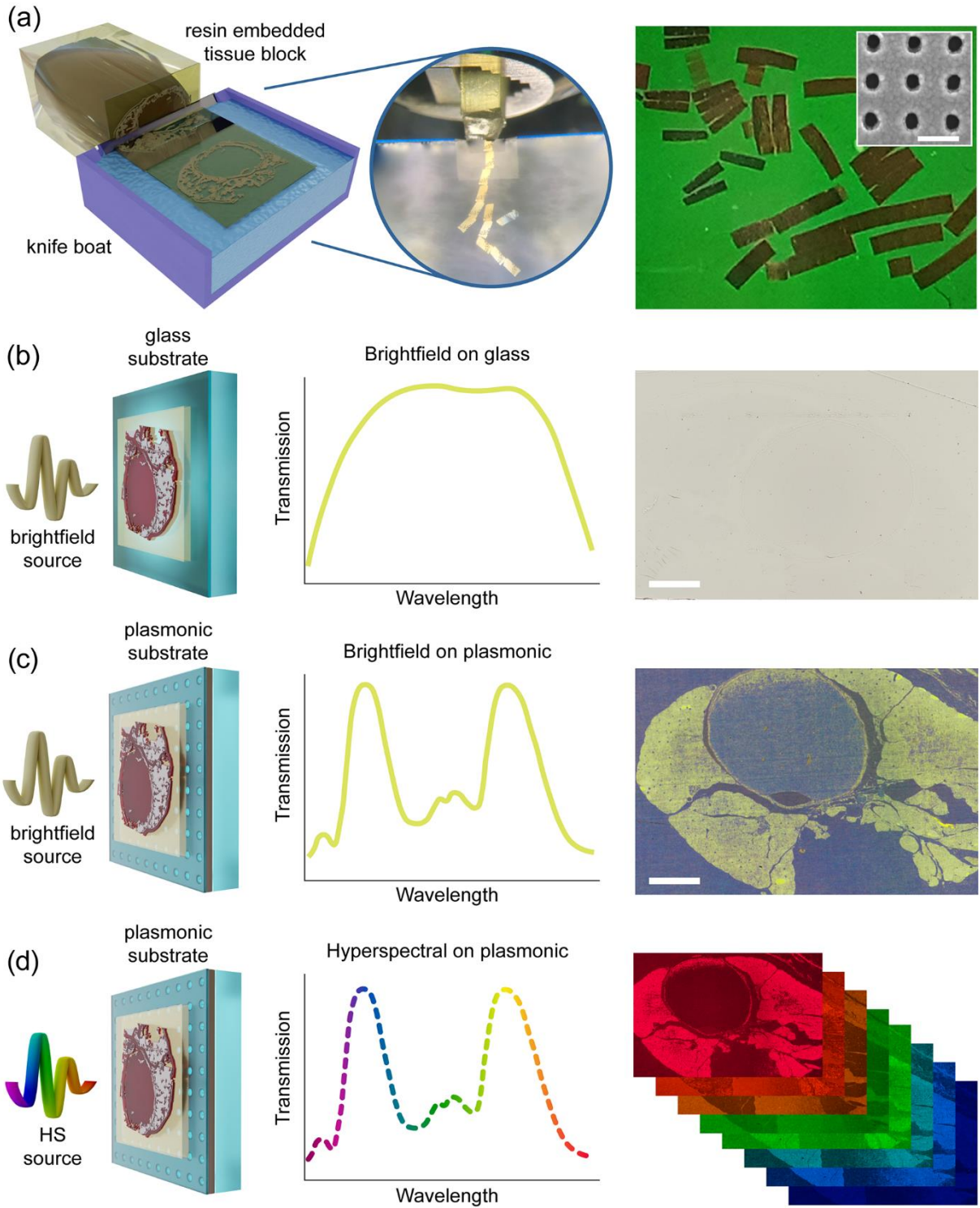

**Figure 1.** (a) Render of hyperspectral imaging workflow used in this study (left). Ultrathin sections of resin embedded tissues are sliced using a diamond knife and collected on the nanoMslide (middle). Photograph depicting optic nerve tissue slices on the slide (right); inset showing an SEM image of the fabricated apertures (scalebar represents 400 nm). Tissue slices are subsequently imaged in different configurations. Schematic representation of the brightfield imaging of slices on glass (b), nanoMslide (c), and hyperspectral imaging of slices on nanoMslide (d). Scalebar represents 100 μm.

## 3. Results and Discussion

As a first demonstration of the capabilities of PE-HSI, a series of oxides (silicon, aluminium, magnesium and tungsten) with refractive indexes of 1.4649, 1.6798, 1.7397 and 1.9866 respectively, were deposited onto a glass and a polarization sensitive nanoMslide device in small discs of ~80 um in diameter. Eight discs of each oxide were deposited with thicknesses of 60 nm for tungsten oxide and 120 nm for all other oxides (**supplementary Figure 1**). These oxides were chosen as their refractive indexes are close, making them suitable to assess the color differentiation and the segmentation process using PE-HSI methods. The samples were subsequently imaged using brightfield imaging and PE-HSI with 111 spectral bands ranging from 460 to 680 nm and 2 nm bandwidths. Imaging was performed using linearly polarized light in TE and TM configurations.

Brightfield images of the discs on nanoMslide show a striking contrast enhancement compared to the ones on glass (**Figure 2a**). The mechanism behind this effect is discussed in depth in our previous studies.[14,17,21] Also, it is notable that the color contrast for TM polarization (polarizer aligned with the 500 nm array periodicity) outperforms the one for TE polarization (polarizer aligned with the 400 nm array periodicity). The associated L, a and b values are depicted in **Table 1**:

**Table 1.** L, a and b values for oxides imaged on nanoMslide at TE (400 nm) and TM (500 nm) polarizations.

|         | TE  |    |     | TM |    |    |
|---------|-----|----|-----|----|----|----|
|         | L   | a  | b   | L  | a  | b  |
| $SiO_2$ | 16  | 24 | 6   | 35 | 29 | 38 |
| $Al_2O_3$ | 29 | 43 | -12 | 42 | 46 | 20 |
| MgO     | 20  | 42 | -16 | 34 | 52 | 10 |
| $WO_3$  | 27  | 40 | 23  | 44 | 49 | 55 |

Although both polarizations performed adequately in terms of the color contrast, the difference in L, a and b values for TM polarization is notably larger for oxides with similar RIs such as Al2O3 and MgO, where the ΔRI is just 0.06. This observation is in line with equation 3, which is based on the dispersion equation 2 applied for two mediums:

$$\Delta\lambda_{sp} = \frac{P\sqrt{\epsilon_m}}{\sqrt{i^2+j^2}}\left(\sqrt{\frac{\epsilon_{d1}}{\epsilon_m+\epsilon_{d1}}} - \sqrt{\frac{\epsilon_{d2}}{\epsilon_m+\epsilon_{d2}}}\right) \qquad (3)$$

Based on this equation, greater color contrast is expected for structures with larger periodicities.

For the HSI feature differentiation, K-means clustering was applied (see Methods). The segmentation results are presented in Figure 2c. For the segmentation, turquoise, blue, green, and red were assigned for the pixels segmented as tungsten, magnesium, aluminium and silicon oxides, respectively. Clearly, the segmentation performed on nanoMslide at both polarizations greatly outperformed the one on glass substrate, where apart from the WO3, the rest of oxides failed to be reliably segmented. The spectra on glass show slight linear variations of the transmitted intensity across the whole spectrum compared to the spectra on nanoMslide, where distinct spectral bands are associated to each oxide (Figure 2b). This effect is mainly attributed to the way the light interacts with the samples. In the case of the brightfield imaging of the glass/sample, the light/matter interaction is described using classical interpretation, whereas in the plasmonic/sample case there is an additional component interpreted by the light interaction with a plasmonic metamaterial.

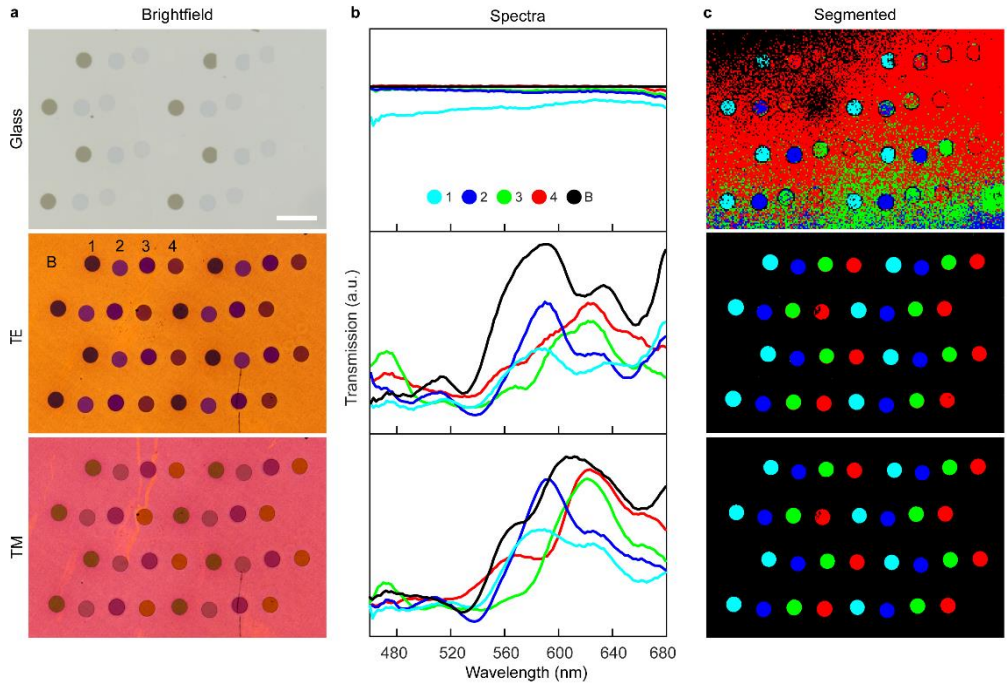

**Figure 2**. Brightfield images (a) and associated HS plots (b) of oxides sample on glass and nanoMslide taken at TE and TM polarizations respectively. c) Segmentation results for glass and nanoMslide at TE and TM polarisations. Turquoise, blue, green, and red were assigned

for pixels segmented as tungsten, magnesium, aluminium, and silicon oxides, respectively. Scalebar represents 200 μm.

As was shown in equation 2, this interaction is highly non-linear, leading to a spectrum composed of a multitude of plasmonic modes (Figure 2b). These modes are strongly dependent on the RI of the sample placed on the plasmonic metamaterial. As a result, a characteristic chromatic contrast is observed for each oxide (Figure 2a). Because of this effect, the segmentation process works more reliably compared to on the glass substrate. Furthermore, compared to the traditional brightfield PE imaging, PE-HSI gives the opportunity to selectively retrieve intensity contrast at specific wavelengths. For instance, it is possible to select wavelengths at which certain features are more highlighted compared to others (**Figure 3**).

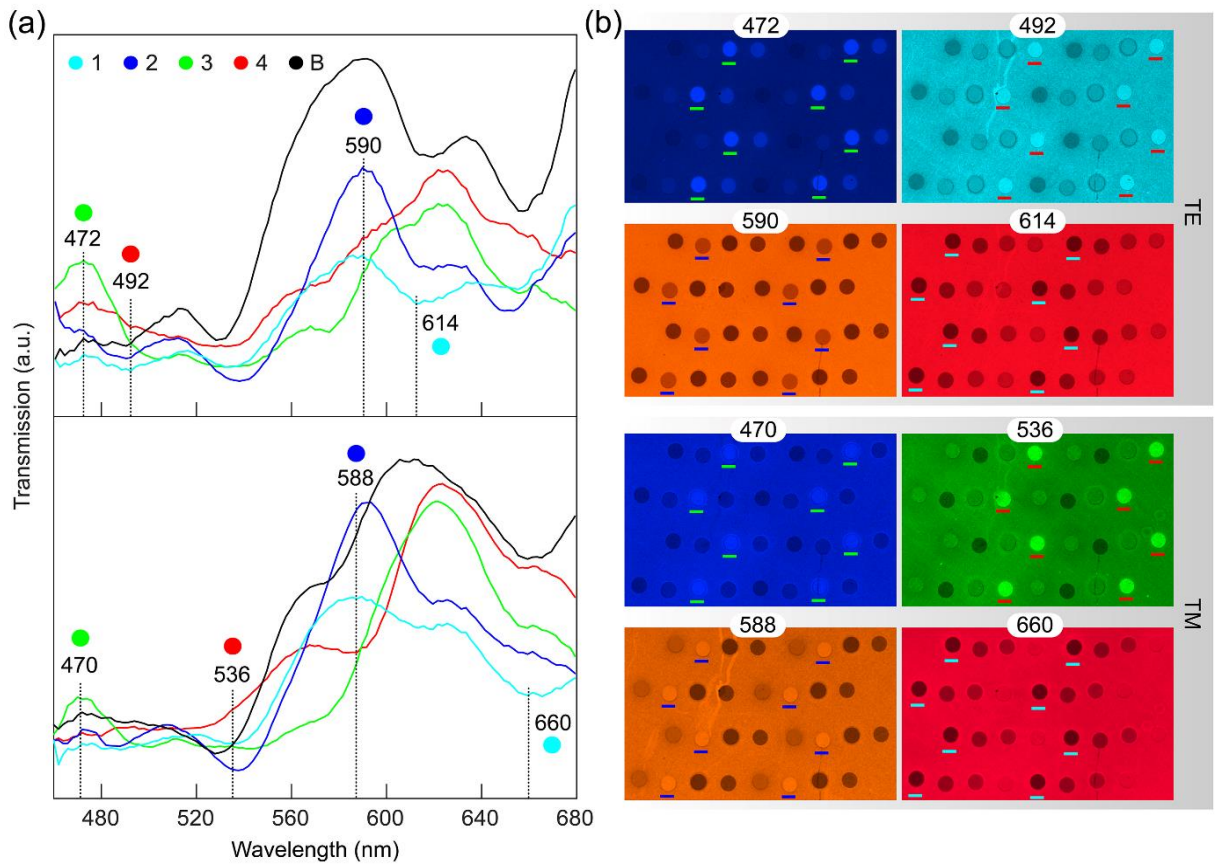

**Figure 3**. (a) HS plots of the oxides sample on nanoMslide taken at TE and TM polarizations respectively. b) Min-max HS images on nanoMslide taken at TE and TM polarisations (corresponding raw images are in the SI). Turquoise, blue, green, and red were assigned to tungsten, magnesium, aluminium, and silicon oxides, respectively. The dashed vertical lines

and the numbers in (a) represent the wavelengths at which the HS images in (b) were taken. Underscored features in (b) are the corresponding oxides. Scalebar represents 200 μm.

For instance, features can be highlighted by the higher brightness levels as features '3', '4' and '2' or lower brightness levels as feature '1'. Such feature discrimination is practically impossible to achieve on glass as demonstrated in Figure 2 a.

## 4. PE-HIS of optic nerve

As it was shown in Figur 1b right, the optical contrast in the biological tissue sections is even more subtle than in the fabricated test sample making the feature recognition and the subsequent segmentation process almost impossible to succeed. However, we demonstrate that using PE-HSI this can be achieved with high contrast even on extremely thin tissue sections (down to 120 nm). **Figure 4a** shows the full hypercube of a mouse optic nerve embedded in resin placed on a polarization sensitive nanoMslide.

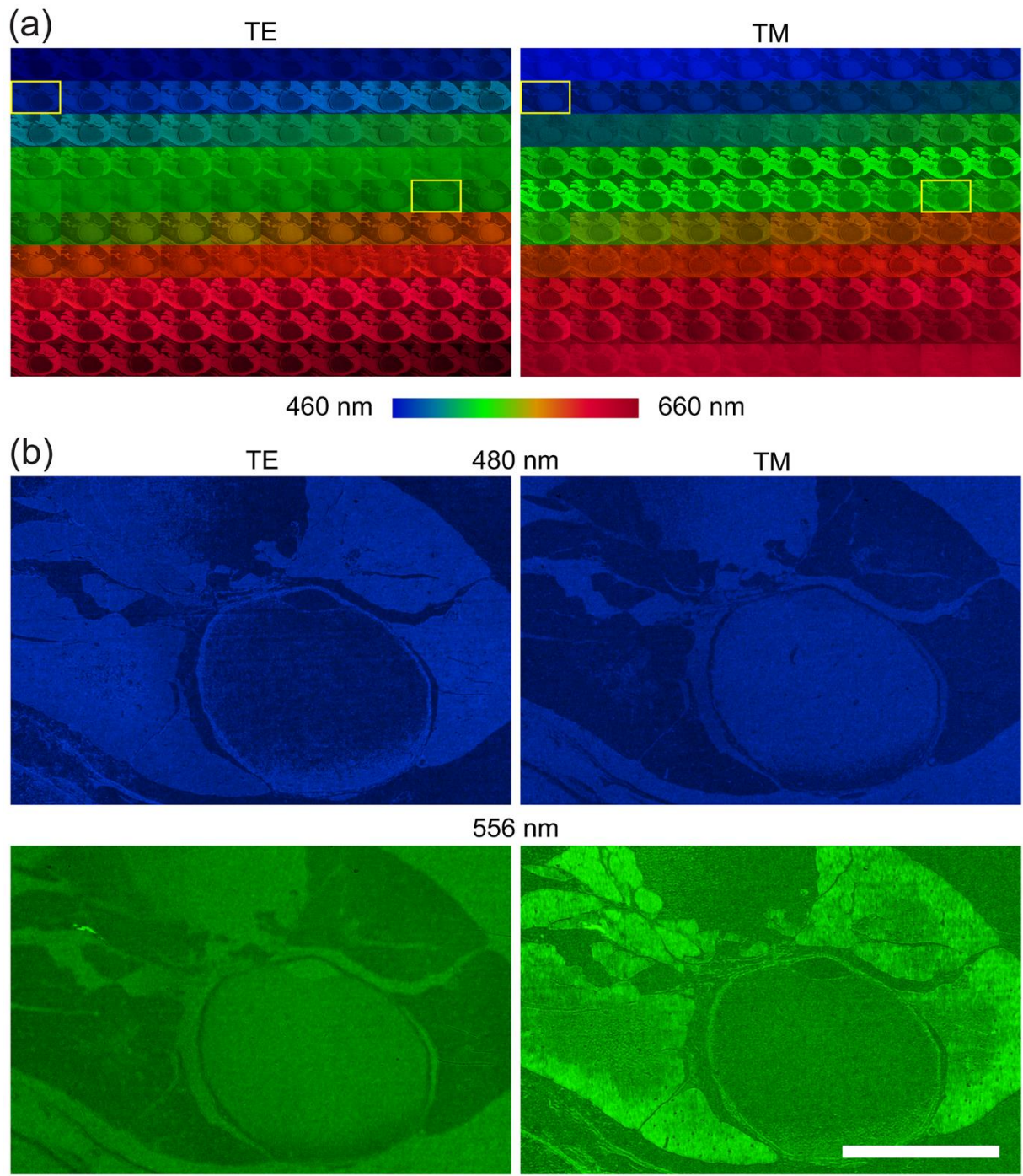

**Figure 4**. PE-HSI of a mouse optic nerve tissue slice (120 nm) embedded in resin imaged at TE and TM polarizations respectively. (a) A complete hypercube taken at 460-660 nm range with 2 nm spectral resolution. (b) Associated HS images taken at 480 and 556 nm wavelengths at both polarizations. The yellow boxes in (a) represent the image position in the stack depicted in (b). Scalebar represents 200 μm.

As in the case of the fabricated sample, the intensity contrast of the optic nerve tissue is greatly improved by placing it on the nanoMslide. This is clearly notable for all wavelength at both polarizations with varying intensity contrast (Figure 4a). An interesting observation is that a contrast reversal can be achieved either at the same wavelength, e.g. 480 nm or 556 nm (Figure 4b), by simply changing the polarization from TE to TM, or at the same polarization (TE or TM) by changing the wavelength from 480 nm to 556 nm. Clearly, certain features appear to be more detectable under a specific wavelength/polarization combination. This is a unique feature characteristic to PE-HSI caused by the highly nonlinear spectral distribution of the plasmonic modes. For this reason, it is superior to brightfield PE imaging.

Next, we applied the segmentation technique used for the oxide sample in order to differentiate features of interest. To improve the lateral resolution down to the nerve cell recognition, we fabricated a 50x50 um array of high-quality circular apertures via focused ion beam lithography (FIB) on which a 75 nm optic nerve section was placed. **Figure 5a** and b show the SEM and the associated optical image of this section placed on the plasmonic array. The segmented images are shown on the right. It is apparent that despite the obvious high-resolution quality of the SEM imaging, the segmentation performed on the SEM image failed to adequately differentiate features of interest mostly due to the lack of chemical sensitivity of the SEM technique. The PE-HSI technique, however, has higher chemical sensitivity compared to the SEM imaging despite been constrained by the diffraction-limited resolution. Features such as axons, Myelin sheath, endoneurium and blood vessels are clearly segmented using the PE-HSI technique. By applying composite imaging technique, which marries the high-resolution of SEM imaging and the high chemical sensitivity of the PE-HSI technique, high-resolution high chemical sensitivity imaging can be achieved (Figure 5b).

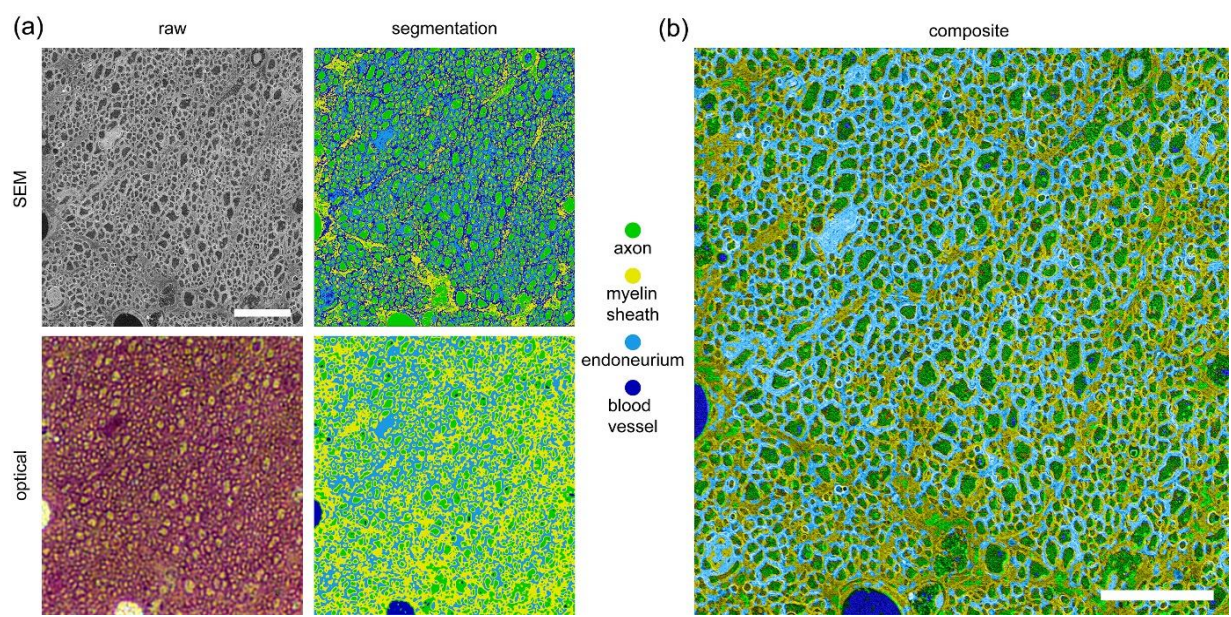

**Figure 5**. Comparison of 75 nm optic nerve tissue section imaged using SEM and PE-HSI. a) shows SEM and optical image of the optical nerve sample with the associated segmentation. b) The composite image based on the SEM raw image and the segmented PE-HSI image. Scalebars represent 10 μm.

## 5. Conclusion

In this work we demonstrated that high compositional sensitivity of plasmon enhanced colorimetric imaging combined with the spectral sensitivity of the HSI led to improved feature recognition and segmentation. By employing well-characterized control samples, features with refractive index difference of just 0.06 were easily distinguishable. Furthermore, it was shown that feature recognition could be performed either by the relative contrast at a specific wavelength or by applying the correlation function using the entire HSI spectrum. Also, polarization sensitive plasmonic devices proved to offer an additional way to expand on the dynamic range of the PE-HSI. These methods were successfully translated in PE-HSI of histological tissue sections, where sections down to 75 nm were segmented for the first time. Additionally, the ability of PE-HSI to complement established techniques such as SEM by providing highly sensitive compositional information has been shown, which open up additional avenues in high-resolution high compositional sensitivity imaging of biological samples for label-free screening.

## 6. Methods

*Device Fabrication:* The periodic arrays of nanoapertures were fabricated using displacement Talbot lithography for large areas (Phabler 100C, Eulitha), or the focused ion beam (FIB) lithography technique (Helios NanoLab 600 Dual Beam FIB-SEM, FEI) for small areas. The devices were fabricated in 150-nm-thick Ag films, using circular nanoapertures of 190 nm in diameter. Polarisation sensitive devices had a 400-nm period in one direction (defined here as the TE direction) and a 500-nm period in the orthogonal direction (defined as the TM direction). FIB fabricated devices had a 450-nm period on a hexagonal pattern. To protect the surface, a very thin layer (10 ± 1 nm) of $SiO_2$ was deposited as the final step after fabrication.

*Oxide sample fabrication:* Four oxides (tungsten, magnesium, aluminium and silicon) were deposited on polarization sensitive devices and on glass substrates by electron-beam evaporation at the rate of 0.5 Å/s (Intlvac Nanochrome II). Tungsten oxide had a final thickness of 60 nm, while the rest of oxides 120 nm respectively.

*Sample Preparation:* Resin embedded optic nerve samples were prepared via fixation using 2.5% glutaraldehyde and post fixed with osmium tetroxide. Serial dehydration using ethanol and acetone was performed prior to infiltration and embedding in epoxy resin. Sections of the nerve were prepared using a Leica UC7 Ultramicrotome.

*HSI imaging:* All images were acquired using a Nikon ti2-e microscope and DS-Ri2 CCD camera. For hyperspectral imaging, a SuperK Compact white laser paired to a SuperK VARIA tuneable bandpass filter was used as a HS source on the Nikon microscope. HSI scans on the oxide sample and large optical nerve sample covered a spectral range from 460 to 660 nm in 2 nm steps using a 5 nm spectral bandwidth, 10 ms exposure time and 4x objective lens. HSI scans of the high-resolution optic nerve sample used a spectra range of 460 to 680 nm in 5 nm steps with a spectral bandwidth of 5 nm, 2 s exposure time and 100x objective lens. Python 3.9.12 was used to control the laser and microscope software using the PyAutoGUI library. The script accounted for focal length variations caused by the changing wavelength by evaluating a focus metric based on a four neighbourhood Laplace operator.[22]

*SEM imaging:* SEM images of the optic nerve sample were captured using a Hitachi SU7000 field emission scanning electron microscope at 1300x magnification. Accelerating voltage was set to 1 kV with a dwell time of 320 µs.

*Image processing and segmentation:* Brightness and contrast adjustments were applied to all brightfield and SEM images in Fiji. The background subtraction function using a 50 pixel radius was applied in Fiji to the hyperspectral images collected on the fabricated oxides on glass to improve segmentation. K-means clustering was performed in MATLAB using the imsegkmeans and kmeans functions for the SEM and hyperspectral data, respectively.

Acknowledgements

This work was performed in part at the Melbourne Centre for Nanofabrication (MCN) in the Victorian Node of the Australian National Fabrication Facility (ANFF). The authors acknowledge the support of the Australian Research Council Discovery Project (DP220103679).

References

[1] G. Lu, B. Fei, *JBO* 2014, *19*, 010901.

[2] D. B. Malkoff, W. R. Oliver, in *Spectral Imaging: Instrumentation, Applications, and Analysis*, SPIE, 2000, pp. 108–116.

[3] A. A. Gowen, C. P. O'Donnell, P. J. Cullen, G. Downey, J. M. Frias, *Trends in Food Science & Technology* 2007, *18*, 590.

[4] R. Rowland, A. Ponticorvo, M. Baldado, G. T. Kennedy, D. M. Burmeister, R. J. Christy, N. P. Bernal, A. J. Durkin, in *Photonics in Dermatology and Plastic Surgery 2019*, SPIE, 2019, pp. 10–20.

[5] J. Stergar, R. Dolenec, N. Kojc, K. Lakota, M. Perše, M. Tomšič, M. Milanic, *Biomed. Opt. Express, BOE* 2020, *11*, 1991.

[6] J. Peller, K. J. Thompson, I. Siddiqui, J. Martinie, D. A. Iannitti, S. R. Trammell, in *Optical Biopsy XV: Toward Real-Time Spectroscopic Imaging and Diagnosis*, SPIE, 2017, pp. 57–68.

[7] Y. Khouj, J. Dawson, J. Coad, L. Vona-Davis, *Frontiers in Oncology* 2018, *8*.

[8] L. Quintana, S. Ortega, R. Leon, H. Fabelo, G. M. Callico, C. Lopez, M. Lejeune, R. Bosch, in *2021 XXXVI Conference on Design of Circuits and Integrated Systems (DCIS)*, 2021, pp. 1–6.

[9] M. Lv, W. Li, R. Tao, N. H. Lovell, Y. Yang, T. Tu, W. Li, *IEEE Journal of Biomedical and Health Informatics* 2021, *25*, 3041.

[10] B. Jansen-Winkeln, M. Barberio, C. Chalopin, K. Schierle, M. Diana, H. Koehler, I. Gockel, M. Maktabi, *Cancers* 2021, *13*, 967.

[11] S. Ortega, H. Fabelo, M. Halicek, R. Camacho, M. de la L. Plaza, G. M. Callicó, B. Fei, *Applied Sciences* 2020, *10*, 4448.

[12] L. Ma, A. Rathgeb, H. Mubarak, M. Tran, B. Fei, *JBO* 2022, *27*, 056502.

[13] S. A. Maier, *Plasmonics: Fundamentals and Applications*, Springer Science & Business Media, 2007.

[14] E. Balaur, C. Sadatnajafi, S. S. Kou, J. Lin, B. Abbey, *Sci Rep* 2016, *6*, 28062.

[15] C. Genet, T. W. Ebbesen, *Nanoscience And Technology: A Collection of Reviews from Nature Journals* 2010, 205.

[16] V. Amendola, R. Pilot, M. Frasconi, O. M. Maragò, M. A. Iatì, *J. Phys.: Condens. Matter* 2017, *29*, 203002.

[17] E. Balaur, S. O' Toole, A. J. Spurling, G. B. Mann, B. Yeo, K. Harvey, C. Sadatnajafi, E. Hanssen, J. Orian, K. A. Nugent, B. S. Parker, B. Abbey, *Nature* 2021, *598*, 65.

[18] S. B. Mahbub, A. Guller, J. M. Campbell, A. G. Anwer, M. E. Gosnell, G. Vesey, E. M. Goldys, *Sci Rep* 2019, *9*, 4398.

[19] D. Zopf, J. Jatschka, A. Dathe, N. Jahr, W. Fritzsche, O. Stranik, *Biosensors and Bioelectronics* 2016, *81*, 287.

[20] M. V. Bashevoy, F. Jonsson, K. F. MacDonald, Y. Chen, N. I. Zheludev, *Opt. Express, OE* 2007, *15*, 11313.

[21] E. Balaur, G. A. Cadenazzi, N. Anthony, A. Spurling, E. Hanssen, J. Orian, K. A. Nugent, B. S. Parker, B. Abbey, *Nat. Photonics* 2021, *15*, 222.

[22] Q. Zhang, Y. Wang, Q. Li, X. Tao, X. Zhou, Y. Zhang, G. Liu, *Journal of Biophotonics* 2022, *15*, e202100366.

Supporting Information

**Figure S1. AFM linescans of oxide sample**

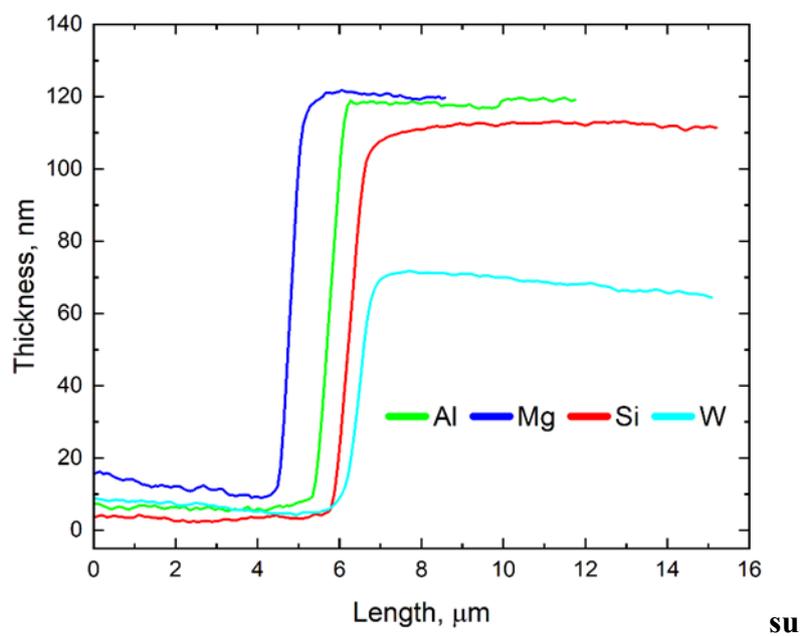

su

**Figure S2. PE-HS images on nanoMslide taken at TE (400 nm) and TM (500 nm) polarisations**

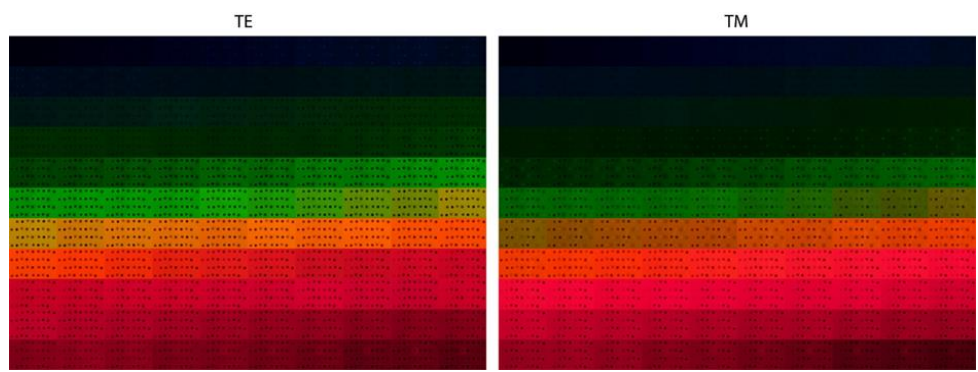